\documentclass[iop,apjl,tighten]{emulateapj}
\slugcomment{ }

\shorttitle{Quantum cyclotron harmonics in X-ray spectra of neutron stars}
\shortauthors{Suleimanov, Pavlov, \& Werner}

\begin{document}

\title{Quantum nature of cyclotron harmonics in thermal spectra of neutron stars
    }


\author{V. F. Suleimanov\altaffilmark{1,2}, G. G. Pavlov\altaffilmark{3}, and K. Werner\altaffilmark{1}}
\affil{}
\email{}


\altaffiltext{1}{Institute for Astronomy and Astrophysics, Kepler Center for Astro and Particle Physics,
   Eberhard Karls University, Sand 1, 72076 T\"ubingen, Germany; suleimanov@astro.uni-tuebingen.de}
\altaffiltext{2}{Dept. of Astronomy, Kazan State University, Kremlevskaya 18, 420008 Kazan, Russia}
\altaffiltext{3}{Pennsylvania State University, 525 Davey Lab., University Park, PA 16802;
%, USA;
pavlov@astro.psu.edu}

\begin{abstract}
Some isolated neutron stars show harmonically spaced absorption features in their
thermal
soft X-ray spectra. 
The interpretation of the features as a cyclotron line
and its harmonics has been suggested, 
but the usual explanation of the harmonics as caused by relativistic
effects 
%in the cyclotron processes 
fails because the relativistic
corrections are extremely small in this case.
We suggest that the 
features
%formed in neutron star atmospheres 
%with quantizing magnetic fields, 
correspond to
the peaks 
in the energy dependence of the free-free opacity in a quantizing magnetic field, known as quantum oscillations.
The peaks
%are associated with opening new channels 
arise when the transitions to
 new Landau levels become allowed with increasing the photon energy; 
%they are 
%particularly high
%for photons polarized across
%the magnetic field.
%These quantum oscillations 
%is an 
%essentially nonrelativistic phenomenon; 
they
are strongly enhanced by
the square-root 
singularities in the phase-space density
of quantum states in the case when the free (non-quantized) motion is
effectively one-dimensional.
%To demonstrate the effects of quantization,
To explore observable properties of these quantum oscillations,
% in the thermal spectra of neutron stars,
we calculate models of
%fully ionized 
hydrogen neutron star atmospheres
with $B \sim 10^{10}$--$10^{11}$ G (i.e., electron cyclotron energy
$E_{c,e}\sim 0.1$--1 keV) and $T_{\rm eff}
 = 1$--3 MK. Such conditions are
thought to be typical for the so-called central compact objects 
in supernova remnants, such as 1E\,1207.4--5209 in PKS 1209--51/52.
We
show that 
observable 
%absorption 
features at the
electron
cyclotron 
%energy and its 
harmonics form at moderately
large values of the 
%effective 
quantization parameter,
$b_{\rm eff}\equiv E_{c,e}/kT_{\rm eff} \simeq 0.5$--20.
The equivalent widths of the features
can reach $\sim 100$--200 eV;
they grow with increasing $b_{\rm eff}$ and are lower for higher harmonics.
%We 
%note that the harmonically spaced absorption features
%detected in some ``dim isolated neutron stars'' 
%could be due to 
%%similar 
%quantum oscillations caused
%by quantization of proton motion in 
%very strong magnetic fields, $B\sim (1$--$3)\times 10^{13}$ G.
\end{abstract}


\section{Introduction}
Observations 
with the X-ray observatories
{\sl Chandra} and {\sl XMM-Newton} have led to the discovery
of absorption features in the thermal spectra of several
isolated (nonaccreting) neutron stars (NSs).
Studying such features, one could infer the
chemical composition, magnetic field, and 
gravitational redshift at the NS surface, but identification of the features
is rather complicated, mostly because of the effects of the
a priori unknown NS magnetic field 
on radiative transitions in the X-ray range.

First 
absorption features in the spectrum of an isolated NS,
centered at about 0.7 and 1.4 keV, 
were discovered
by Sanwal et al.\ (2002) 
in {\sl Chandra} observations of
1E\,1207.4--5209
(hereafter 1E\,1207), 
which is the central compact object (CCO) of the supernova
remnant (SNR) PKS 1209--51/52.
Similar to other CCOs (Pavlov et al.\ 2002, 2004;
de Luca 2008; Halpern \& Gotthelf 2010), 
1E\,1207 does not show the usual pulsar activity, such as radio and $\gamma$-ray
emission or a pulsar wind nebula.
This NS has the period $P=0.424$ s
(Zavlin et al.\ 2000) and shows a thermal-like
X-ray spectrum 
($T_{\rm eff} \sim 1$--3 MK, depending on the model;
Zavlin et al.\ 1998).
Further observations with {\sl XMM-Newton} suggested the presence of
two more features, at 2.1 and 2.8 keV (Bignami et al.\ 2003).

 As the energies of the spectral features are harmonically spaced, it seems natural to
interpret them 
as (gravitationally redshifted) electron cyclotron line
and its harmonics (e.g., due to the
 ``resonance cyclotron scattering'', by analogy with accreting
X-ray pulsars; Bignami et al.\ 2003), formed in the
magnetic field $B = 6\times 10^{10} (E_{c,e}^\infty/0.7\,{\rm keV}) (1+z)$ G,
where 
$E_{c,e}^\infty (1+z) = E_{c,e}=\hbar eB/m_ec$ is the electron cyclotron 
energy, 
and $z$ is
the gravitational redshift.
This estimate for the magnetic field is consistent with
the upper limit
$B_{sd}< 3.3\times 10^{11}$ G obtained by Gotthelf \& Halpern (2007) from
the upper limit on the pulsar's 
period derivative $\dot{P}$.

However, 
the cyclotron harmonic interpretation encounters a serious problem 
because 
the ratio of the 
emissivities (or oscillator strengths) of consecutive harmonics
is known to be proportional to the parameter
 $\xi_e\equiv {\rm max}(kT, E_{c,e})/m_ec^2$ (see, e.g., Pavlov
et al.\ 1980a,b), i.e., the emission and absorption in cyclotron harmonics
is an essentially relativistic phenomenon. 
As this parameter is quite small for 1E\,1207, 
$\xi_e
\sim 10^{-3}$, 
observable cyclotron harmonics can hardly be expected, even with allowance
for the fact that the strength of 
a feature
formed in an optically thick medium depends 
on the oscillator strength nonlinearly, and it also depends on other
factors, such as the temperature gradient at the depth where the
feature is formed. 
We should note that the conditions in the relatively cold, low-field
atmospheres of 1E\,1207 and other CCOs
 (where the free-free absorption and emission dominate over the electron 
scattering of photons) are quite different from those
in accreting X-ray pulsars,
where the parameter $\xi_e$ is a factor of 30--100 larger,
which explains the observability of electron cyclotron harmonics in those
objects (e.g., two harmonics, in addition to the fundamental at 26 keV,
 have been observed in the binary pulsar V0332+53; Pottschmidt
et al.\ 2005) .

%Other objects in which harmonically spaced absorption features
%have been
%apparently observed in the soft X-ray range
%($\sim 0.1$--1 keV) are the so-called 
%``dim isolated NSs'' (DINSs; see, e.g., Kaplan 2008). 
%These objects are also radio-quiet,
%and their spectra are also thermal-like, with typical temperatures
%$T\sim 0.5$--1 MK, but their magnetic fields, 
%$B_{sd}\sim (1$--$3)\times 10^{13}$ G,
%estimated from the spindown rates,
%are much higher than those of CCOs.
%Therefore, it was suggested (e.g., Haberl 2007) that the features are the
%proton cyclotron line at $E=E_{c,p}=(m_e/m_p)E_{c,e}=
%0.63 (B/10^{14}\,{\rm G})$ keV
% and its harmonics. This interpretation, however,
%looks 
%quite implausible given the extremely small 
%value of the parameter $\xi_p\equiv 
%{\rm max}(kT, E_{c,p})/m_pc^2 < 10^{-6}$
%that determines the relative strengths of consecutive proton cyclotron
%harmonics. 
%Thus, 
%the harmonically spaced absorption features
%in thermal spectra of isolated NSs
%(CCOs and DINSs) cannot be 
%caused by the ``usual'' cyclotron processes.

Resonances at cyclotron harmonics (also known as quantum oscillations)
 can, however, arise even in 
a nonrelativistic plasma due to
quantization of rotation of a charged particle in a magnetic field.
In particular, the energy of, e.g., a nonrelativistic electron 
is 
\begin{equation}
\epsilon_n(p_z) = \left(n+\frac{1}{2}\right) E_{c,e} +\frac{p_z^2}{2m_e}\,,
\end{equation}
 where $n=0$, 1, 2, $\ldots$
numerates the Landau levels, and $p_z$ is the 
(continuous) electron's momentum along
the magnetic field.
As a result of the quantization,
various properties of matter 
show discontinuities when 
the number of the involved Landau levels changes (e.g., when a transition
to a higher Landau level becomes allowed with increasing the energy
of the photon that causes this transition). Moreover, because the electron
motion is continuous only in one dimension, 
the phase space density of
final electron states in such transitions experiences square-root 
singularities at 
these discontinuities, which gives rise to 
logarithmically high resonances at harmonics of the 
cyclotron energy
(see Section 2 for more detail). 

Such quantum resonances in free-free absorption of photons in a magnetic
field were first noticed by Pavlov \& Panov (1976; hereafter PP76) 
who derived the
spectral opacities and emissivities of a plasma with $kT \ll m_ec^2$
for different photon polarizations. As the free-free transitions are
an important source of opacity in NS atmospheres, one can expect
the resonances to manifest themselves as
absorption spectral features in thermal spectra of NSs.
This was qualitatively demonstrated by Pavlov \& Shibanov (1978) 
who considered emission from an 
``atmosphere'' in which the
source
function in the radiative transfer equation
 was approximated by a linear function of
optical depth.
More realistic, self-consistent NS atmosphere models have been 
developed later, starting from Shibanov et al.\ (1992), but they used
approximate formulae for the free-free opacity, in which the quantum
resonances were neglected. 

%In this {\it Letter} we 
Here we present first calculations of 
realistic magnetized NS atmosphere models with the 
quantum resonances at electron cyclotron harmonics in the free-free opacity
taken into consideration and show that harmonically spaced absorption features are
significant in the emergent spectra. These calculations demonstrate that
thermal spectra of NSs with a surface magnetic field $\sim 10^{10}$--$10^{11}$ G can
exhibit several such features in the observable X-ray range, and, in 
particular, the features observed in the spectrum of 1E1207 can be
interpreted as due to the quantum effects.

\section{
Quantum oscillations of opacities}

We consider hydrogen NS
atmospheres with magnetic fields $\sim 10^{10}$--$10^{11}$ G,
such that the electron cyclotron energy, $E_{c,e}\sim 0.1$--1 keV, and its
first several harmonics are in the soft
X-ray range, observable with {\sl Chandra} and {\sl XMM-Newton}.
We assume that the atmosphere is hot enough,
$T_{\rm eff} \gtrsim 10^6$ K, to neglect the small fraction
of neutral atoms.
The opacity in such a fully ionized
 atmosphere is determined by two processes:
free-free absorption (inverse bremsstrahlung) and 
scattering of photons by electrons\footnote{Scattering by protons,
as well as the proton contribution into the free-free absorption,
 can be neglected at
the relatively low magnetic fields, when $E_{c,p}\ll E$.},
modified by the magnetic
field. 
We assume that the atmosphere is cold enough, $T_{\rm eff} \lesssim 10^7$ K, to
neglect the change of photon energy in the scattering
process (comptonization effects).
Moreover, we calculate the opacities in the so-called ``cold plasma approximation''
(Ginzburg 1970), which implies
that 
relativistic effects,
such as the cyclotron emission (hence also absorption
and scattering) at the harmonics
of the electron
cyclotron energy, can be neglected, as well as
 the thermal motion effects, including the Doppler broadening of
the cyclotron resonance.

The transfer of high energy radiation in a magnetized plasma is described
by two coupled transfer equations for the ordinary and
extraordinary  normal modes (Gnedin \& Pavlov 1974).
The normal mode polarizations and opacities 
depend not only on the photon energy $E$ but also
 on the angle $\theta_B$ between the wave vector
and the magnetic field; 
they can be written as (e.g., Pavlov et al.\ 1995)
\begin{eqnarray} \label{u1}
    k_{1,2} & = &
%(E,\theta_{\rm B}) =
\sum_{\alpha=-1}^{+1} \frac{E^2}{(E+\alpha E_{c,e})^2 + (\gamma_r + \gamma_\alpha)^2}\times
%E_{\rm cyc})^2}
%~\zeta_{\alpha}~
\\ \nonumber
&& \left[k_{\rm T} +
%\kappa_\alpha^{\rm ff}(E)
k_{\rm cl}^{\rm ff}(E) g_\alpha(E)\right]\,|e^{1,2}_{\alpha}(E,\theta_{\rm B})|^2,
\end{eqnarray}
where 
$k^{\rm ff}_{\rm cl}(E)$ and $k_{\rm T}$ are the 
``classical''
 free-free and Thomson opacities at $B=0$,
$g_{\pm 1}(E) = g_\perp(E)$ and $g_0=g_\|(E)$ are the Gaunt factors for radiation 
polarized across and along the magnetic field, 
$\gamma_r=(2/3)E^2(e^2/\hbar m_e c^3)$ 
and $\gamma_\alpha = \gamma_r\, [k^{\rm ff}_{\rm cl}(E)/k_{\rm T}]\, g_\alpha(E)$ are
the radiative and collisional widths, and 
$e_\alpha^{1,2}(E,\theta_B)$ 
are the projections
 of the normal mode polarization vectors (see Kaminker et al.\ 1982, 1983
for a detailed description of the normal mode polarizations and opacities
in the cold plasma approximation). 

Various forms for the
 magnetic Gaunt factors $g_{\perp,\|}(E)$  
were derived 
by 
PP76\footnote{PP76 used
 Coulomb logarithms,
$\Lambda_{\perp,\|}=(\pi/\sqrt{3})\, g_{\perp,\|}$, instead
of the Gaunt factors.}
who calculated the dependence of the Gaunt factors on the
dimensionless photon energy $x=E/kT$ for several values
of the 
``quantization 
 parameter'' $b_e=E_{c,e}/kT$ and obtained simple expressions for
limiting cases.
Examples of the energy dependence of the Gaunt factors are shown in Figure 1,
for $b_e=5$, 0.05, and 0. 
In our calculations we used
Equations (30) and (31) from PP76, 
\vskip 0.1cm
\begin{eqnarray} \label{u2}
   g_{\|} & = & \frac{3\sqrt{3}}{2\pi} b_e^{1/2} e^{x/2}
  \int_0^{\infty}
\cos{xt}\,\times    \\ \nonumber
&& \left[\frac{\sqrt{A}}{A-D}-
 \frac{D}{(A-D)^{3/2}}
\ln{\frac{A^{1/2}+(A-D)^{1/2}}{D^{1/2}}} \right] dt, \\
g_{\bot} & = & \frac{3\sqrt{3}}{4\pi} b_e^{1/2} e^{x/2} \int_0^{\infty}
\cos{xt}\,\times
   \\ \nonumber
&& \left[-\frac{\sqrt{A}}{A-D}+
  \frac{2A-D}{(A-D)^{3/2}}
\ln{\frac{A^{1/2}+(A-D)^{1/2}}{D^{1/2}}} \right] dt,
\end{eqnarray}
where
\begin{eqnarray} \label{u4}
  &&  A =  b_e \left(t^2 + \frac{1}{4}\right), ~~\\ \nonumber
   &&   D  =  \coth\frac{b_e}{2} - \frac{\cos(b_et)}{\sinh(b_e/2)} + \frac{m_{\rm e}}{m_{\rm p}}\left(t^2+\frac{1}{4}\right).
\end{eqnarray}
Figure 1 demonstrates that the energy dependences of the
magnetic Gaunt factors 
oscillate around the smooth non-magnetic curve,
with maxima at the cyclotron energy and its harmonics. The oscillations are
strong in $g_\perp(E)$, especially at larger values of $b_e$, 
being almost imperceptible in $g_\|(E)$. 
If the last term in $D$ is omitted 
(i.e., terms $\propto m_e/m_p$ are neglected),
the cyclotron resonances in $g_\perp(E)$ diverge
logarithmically (PP76; see the dashed curves in the upper panel of Figure 1).
The origin of the 
oscillations and divergences
can be 
understood as follows.

 For an electron transition
$n,p_z \to n',p_z'$ with absorption of a photon with
energy $E$, the conservation of energy is described by the delta-function:
\begin{eqnarray}
&& \delta\left[
\frac{p_{z}'^2-p_{z}^2}{2m_e} +sE_{c,e} - E\right]
 \\ \nonumber
&& = \frac{m_e}{P}\left[\delta(p_z'-P) + \delta(p_z'+P)\right]
\Theta(P^2) ,
\end{eqnarray}
where $P=[2m_e (E-sE_{c,e} +p_z^2/2m_e)]^{1/2}$, $s=n'-n$, and
$\Theta(P^2)=
\Theta(E - sE_{c,e} + p_z^2/2m_e)$ is the unit step function.
The equation for the transition rate contains integrations over 
$p_z'$ and $p_z$ and summations over $n$ and $n'$ (or over $n$ and $s$).
Because of the step function in Equation (6), 
a new term 
in the sum over $s$ is added whenever the increasing $E$ passes through 
$kE_{c,e}-p_z^2/2m_e$ ($k=1,2,\ldots$). 
Each of the added terms contains a singular factor
$\propto (E-kE_{c,e} + p_z^2/2m_e)^{-1/2}$ at $E\to kE_{c,e}-p_z^2/2m_e +0$.
Further averaging over $p_z$ with the one-dimensional
electron distribution function turns the square-root
singularities into logarithmic singularities,
$\propto \ln|E-kE_{c,e}|^{-1}$.
Such behavior is seen in $g_\perp(E)$ but not in $g_\|(E)$ because
in $g_\|(E)$ the
singularity is canceled by the zero of the 
square of the transition matrix element for the longitudinal
polarization [$\propto (p_z'-p_z)^2$;
see Equations (20) and (26) in PP76].

\begin{figure}
\begin{center}
\vskip -0.5cm
\includegraphics[angle=0,width=8.3cm]{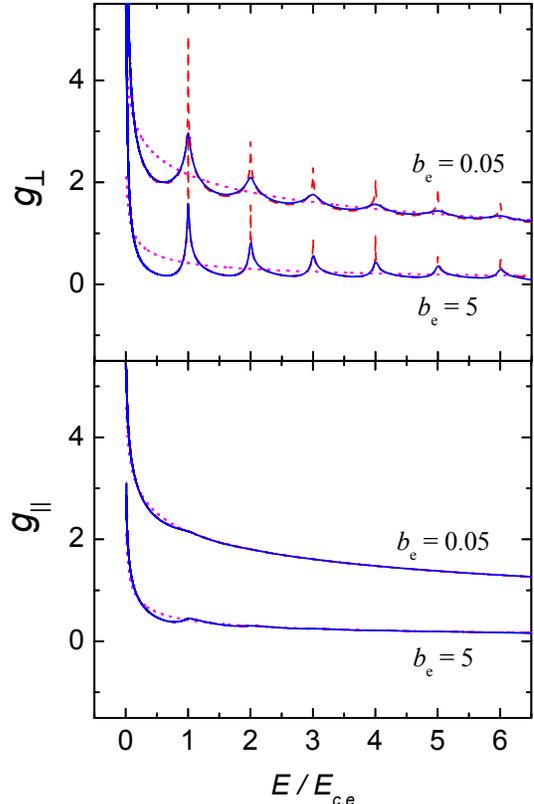}
\vskip -0.8cm
\caption{\label{fig1}
Dependence of the transverse and longitudinal
Gaunt factors on relative
photon energy for an electron-proton plasma with
$b_e\equiv
E_{c,e}/kT$ = 5 and 0.05.
The solid curves show
the Gaunt factors
with proton motion and recoil taken into account, the dashed curves show
the Gaunt factors in the limit of $m_p\to \infty$, and
the dotted curves are
the Gaunt factors at $B=0$.}
\end{center}
\end{figure}

There exist a number of mechanisms that broaden the peaks $E=kE_{c,e}$
and remove the logarithmic singularities (PP76). The last term in $D$
(Equation [5]) accounts for one of such mechanisms, 
associated with the proton 
recoil (the broadened peaks
of finite heights are shown
by solid lines in Figure 1). When the magnetic field decreases, the
distances between the peaks
eventually become smaller than their widths, 
the oscillations disappear, and the magnetic Gaunt factors turn into the 
Gaunt factor at $B=0$. 

We emphasize that the oscillations in $g_{\perp,\|}(E)$ are solely due
to quantization of the electron motion across the magnetic field,
because of which 
the motion in only one dimension
(along 
the magnetic field) remains non-quantized and the square-root singularities
appear after the integration over the final longitudinal momentum $p_z'$
(see Equation [6]).  
At $B \to 0$, when the quantization disappears, 
the integration over the final (three-dimensional)
momentum with allowance for the energy conservation,
$\int \ldots \delta[(p'^2-p^2)/2m_e -E]\, d^3p' = 
4\pi m_e P \int \ldots dp'$
[where $p'$ and $p$ are the moduli of the final and initial momentum vectors,
and $P=(p^2+2m_eE)^{1/2}$],
does not lead to singularities or oscillations. 

\section{Atmosphere models}

To calculate the 
magnetic NS 
hydrogen atmosphere
models, we use our recently
developed code
\citep{sul:09}, 
assuming the magnetic field normal to the 
stellar surface. 
We assume full ionization and neglect the vacuum
polarization by the magnetic field 
because  relatively weak 
magnetic fields are considered. 
For the opacities, we use the equations by
\cite{vAL:06}, substituting there the above-described magnetic 
Gaunt factors\footnote{\cite{vAL:06} use the magnetic Gaunt factors by 
Nagel (1980), which are applicable only at $kT\ll E_{c,e}$ and
$E<E_{c,e}-O(kT)$
(see the Appendix in Kaminker et al.\ 1983).}; 
these equations reduce to Equation (2) when the proton contribution is negligible.

\begin{figure}
\begin{center}
\vskip -0.4cm
\includegraphics[angle=0,width=9.0cm]{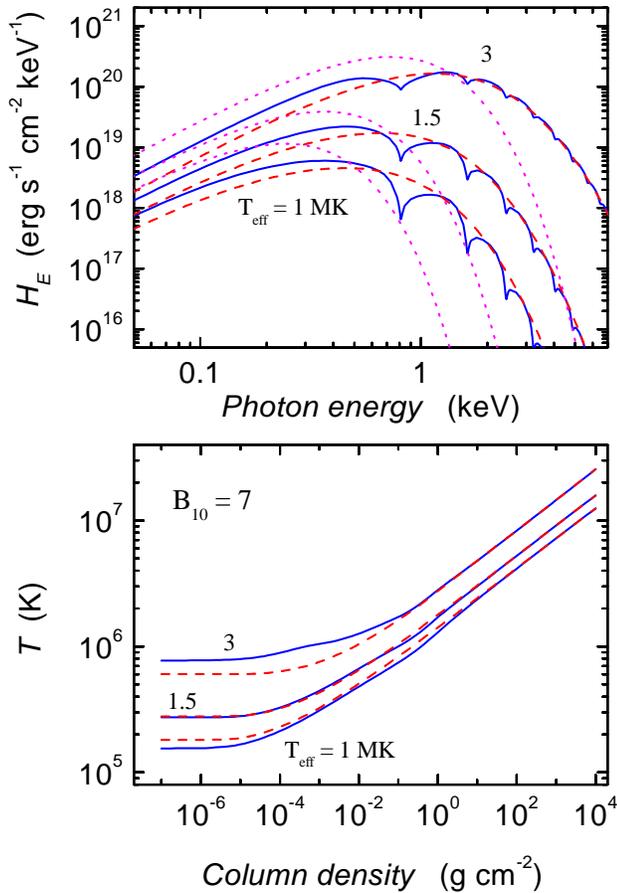}
\vskip -0.7cm
\caption{\label{fig2}
Emergent spectra ({\it top}) and temperature structures ({\it bottom}) for
NS atmospheres with
the magnetic field $B=7\times 10^{10}$ G (solid curves) and $B=0$ (dashed
curves) for three
effective temperatures, $T_{\rm eff}=1$, 1.5, and 3 MK.
The dotted curves in the top panel show the blackbody spectra for the same temperatures.}
\end{center}
\end{figure}

Assuming the surface gravitational
acceleration of $1\times 10^{14}$ cm s$^{-2}$, we
calculated two sets of the atmosphere models.
In the first set the
magnetic field strength was fixed ($B = 7 \times 10^{10}$ G) 
and models with three different 
effective temperatures ($T_{\rm eff}=1$, 1.5 and 3 MK) were constructed. 
A common property
of the emergent spectra, as shown in
Figure \ref{fig2} (top panel),
is prominent
absorption features
at the cyclotron energy
and its harmonics, $E_k = kE_{c,e} = 0.81 k$ keV ($k=1,2,\ldots$), 
which originate from the
quantum peaks in $g_\perp(E)$.
The equivalent width $W_k$  of the $k$-th feature decreases with increasing $k$
(similar to the strengths of the quantum peaks in the transverse Gaunt factor)
as well as with increasing $T_{\rm eff}$
(i.e., with decreasing the effective
quantization parameter, $b_{\rm eff}\equiv E_{c,e}/kT_{\rm eff}
\approx 9.4$, 6.3, 3.1 for the three models). 
For instance, 
$W_1= 240$, 200 and 180 eV at $T_{\rm eff}=1$, 1.5 and 3 MK, respectively,
while $W_2 = 190$, 140 and 55 eV at the same temperatures. 

In the 
top panel of Figure \ref{fig2}
we also show the emergent spectra for 
non-magnetic atmosphere models with the same effective temperatures. 
These spectra 
are close to the ``continua'' of the
corresponding  magnetic spectra at energies higher than the cyclotron
energy,  but  
they lie below the magnetic spectra at lower energies as
the radiation emerges from deeper and hotter layers
 in the magnetic case  because of lower
opacities 
 at $E \ll  E_{c,e}$.

\begin{figure}
\begin{center}
\vskip -0.4cm
\includegraphics[angle=0,width=9.0cm]{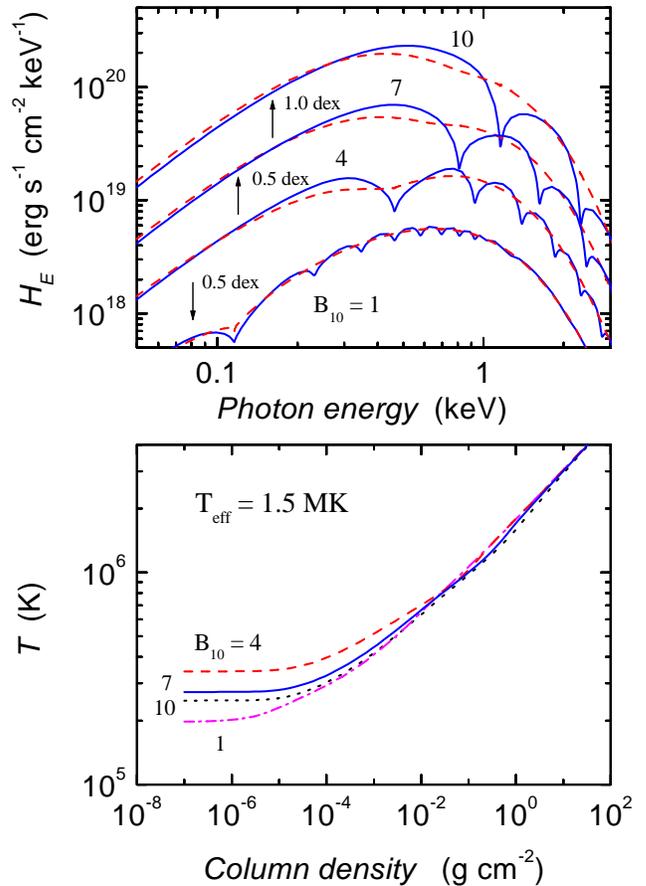}
\vskip -0.7cm
\caption{\label{fig3}
{\it Top panel:} Emergent spectra
for magnetic
NS atmospheres with $T_{\rm eff}=1.5$ MK and
 different magnetic fields ($B=1$, 4,  7 and 10 $\times 10^{10}$ G),
calculated with magnetic and non-magnetic Gaunt factors.
For clarity, the spectra for $B_{10}= 10$, 7 and 1 are shifted
along the ordinate axis by factors $10^{+1}$, $10^{+0.5}$
and $10^{-0.5}$.
{\it Bottom panel:} Temperature structures of the models
 with different magnetic fields: 1 (dash-dotted curve), 4 (dashed curve),
7 (solid curve), and 10 $\times 10^{10}$ G (dotted curve).}
\end{center}
\end{figure}

The second set consists of four models with different magnetic field
strengths ($B=1$, 4, 7, and 10 $\times 10^{10}$ G) at the same
 effective temperature 
($T_{\rm eff} =$ 1.5 MK).
The emergent spectra 
are shown in 
the top panel of Figure \ref{fig3}. 
Again, the equivalent widths of the absorption features decrease with increasing harmonic number $k$ and decreasing effective quantization parameter
($b_{\rm eff}=0.9$, 3.6, 6.3 and 9.0 for the four models).

The radiation spectra  in the 
ordinary (O) and extraordinary (X) polarization modes
are shown in the top panel of Figure 4,
together with the total (X+O)
spectrum, for the model
with 
$T_{\rm eff} = 1.5$ MK and $B = 7 \times 10^{10}$ G. 
As the X-mode opacity  
is low at $E \ll E_{c,e}$
due to the factor $E^2/[(E\pm E_{c,e})^2+\gamma^2]\approx (E/E_{c,e})^2\ll 1$,
the X-mode radiation emerges from deep, hot 
layers and dominates at these energies.
But at 
the cyclotron 
resonance the opacity in the X-mode is strongly increased,
and the 
 total spectrum is mainly determined by the O-mode, whose opacity 
is about the same as  at $B=0$.
As a result, the cyclotron absorption feature is 
very strong in 
the X-mode spectrum but 
not in the total spectrum because the X-mode flux is small
(Pavlov \& Shibanov 1978).
This explains the lack of a strong cyclotron feature in the atmosphere
spectra calculated with the non-magnetic Gaunt factors (see Figure 3, top
panel). Instead of a strong absorption line, we see in these spectra
a broad, shallow depression at $E\lesssim E_{c,e}$ (in the red wing of
the cyclotron line), where the X-mode starts to dominate in the
emergent spectrum.
In contrast, the 
transverse Gaunt factor is present in the opacities of both modes,
so that the corresponding absorption features are prominent in the
outgoing spectrum. 
 
\begin{figure}
\vskip -0.4cm
\includegraphics[angle=0,width=9.0cm]{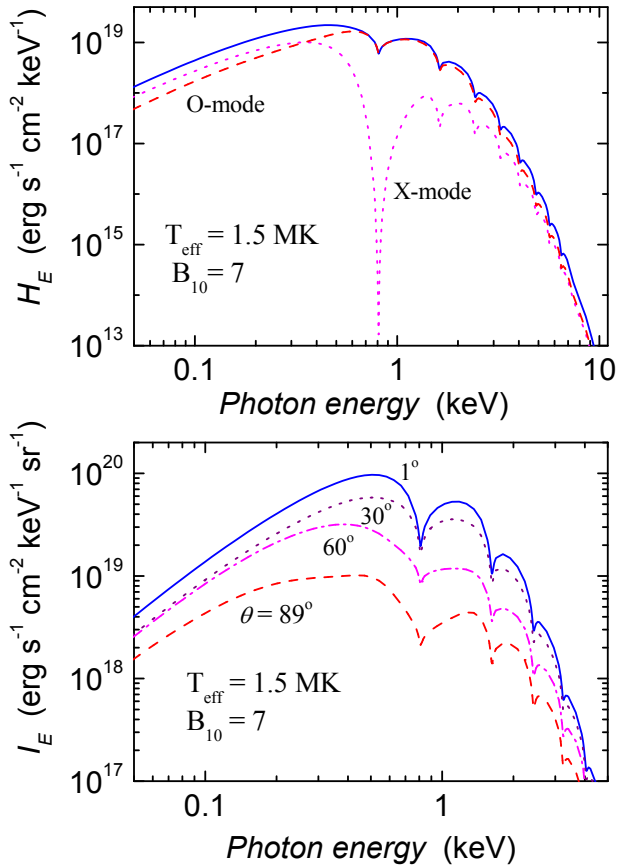}
\vskip -0.7cm
\caption{\label{fig4}
{\it Top panel:} Emergent spectrum of the
magnetic
NS atmosphere with
$B=7\times 10^{10}$ G and
$T_{\rm eff}=1.5$ MK (solid curve) together with emergent spectra in ordinary
(O-mode, dashed curve) and extraordinary (X-mode, dotted curve) modes.
{\it Bottom panel:}
Spectra of the emergent specific intensity
for the same model at different
angles to the surface normal (indicated near the curves).}
\end{figure}
   
Spectra of emergent specific intensity 
for the model with  $T_{\rm eff} =$ 1.5 MK and 
$B = 7 \times 10^{10}$ G for various angles $\theta$
between the line of sight and the surface normal
are shown in the bottom panel of Figure \ref{fig4}.
The quantum oscillations in the intensity spectra are seen at any
angle; the equivalent width of the corresponding absorption
features depends on $\theta$ nonmonotonously
(e.g., $W_1 \simeq 220$, 160, 320 and 330 eV for $\theta=1^\circ$, 
$30^\circ$, $60^\circ$ and $89^\circ$, respectively).
From the comparison of the spectra at different $\theta$ we see that
the angular distribution of the emergent radiation in the absorption
features is different from that in the continuum. For instance,  
the specific intensities are more peaked towards
the surface normal at energies outside the absorption features, 
while the radiation is generally more isotropic at the center of the
features.

The bottom panels of Figures 2 and 3 show the temperature structures for
the six atmosphere models. The temperature structure is only
slightly affected by the relatively low magnetic fields. In particular,
the magnetic field raises
the temperature of upper atmospheric layers
at $3\lesssim b_{\rm eff} \lesssim 6$,
apparently because of cyclotron heating, while outside of this range
 the surface temperature becomes lower.

\section{Discussion}

We have presented
computations of fully ionized hydrogen atmospheres of 
NSs
for magnetic fields $B \sim 10^{10}$--$10^{11}$ G, such
that the electron cyclotron energy, $E_{c,e}\sim 0.1$--1 keV,
is within the range of energies where the thermal emission from isolated
NSs
is usually observed.
We have shown that the 
peaks in the energy dependence of the
free-free opacity, caused by 
the quantization of the
electron rotation around the magnetic force lines, 
lead to absorption features (quantum oscillations) at the electron
cyclotron energy and its harmonics in the atmosphere
spectra. 
The quantum oscillations are best observable at moderately large values of the
quantization parameter,
$0.5 \lesssim b_{\rm eff}
\lesssim 20$,
when the quantization is significant but the features are not too far in the Wien tail
of the spectrum.
The equivalent widths of the absorption features
reach $\sim 100$--200 eV in the examples considered; they grow with increasing $b_{\rm eff}$
and are lower for higher harmonics.

At least two absorption features, at energies 0.7 and 1.4 keV,  
%with  similar
with equivalent widths $\sim 100$ eV have been observed in the X-ray spectrum of the CCO 
1E\,1207 (Sanwal et al.\ 2002). Since
the thermal and cyclotron energies are essentially nonrelativistic for this object
($kT_{\rm eff}/m_ec^2 < E_{c,e}/m_ec^2 \sim 10^{-3}$),
the 
features cannot be explained as due to the
commonly known cyclotron processes. On the other hand, as 
$b_{\rm eff}\sim 3$--5, the features can be naturally interpreted as
 caused by the quantum oscillations.
We note that our calculations
support the reality of the suspected features at 2.1 and 2.8 keV in the
1E\,1207 spectrum
\citep{big:03}.
 To confirm 
our interpretation and infer the properties of the 
1E\,1207's atmosphere,
 the phase-dependent spectra of 1E\,1207 should be compared with
the model spectra obtained by integrating the
specific intensities over the visible NS surface at different rotation
phases, for various orientations of the rotation and magnetic axes.

% should be an accurate phase-resolved analysis of
%the 1E\,1207's spectrum with the aid of our models
%must be performed. 

In addition to 1E\,1207, there exist other 
NSs with similar magnetic fields
and temperatures whose emission can be described by our atmosphere models.
In particular, other CCOs, which likely have magnetic fields $\sim 10^{10}$--$10^{11}$ G
(Halpern \& Gotthelf 2010, and references therein), are expected to show quantum 
oscillations in their soft X-ray spectra. 
However, conditions for their observation
may be less favorable.
For example, 
for the spin-down magnetic field $B_{\rm sd}
=3.1 \times 10^{10}$ G of 
the CCO J1852+0040 in the Kesteven 79 SNR  
(Halpern \& Gotthelf 2010),
the spectrum 
should show absorption features at the
cyclotron energy,
$E_{c,e}^\infty =0.36 (1+z)^{-1}$ keV
and its harmonics.
However, 
as the spectrum below
$\sim 1$ keV is strongly absorbed by
the ISM ($N_{\rm H}\approx 1.8\times 10^{22}$ cm$^{-2}$), only high harmonics, 
whose equivalent widths are small, might be detectable. Interestingly, the spectrum of
this CCO, detected by the {\sl XMM-Newton} EPIC detectors
  (see Figure 3 in Halpern \& Gotthelf 2010), shows a hint
of an absorption feature at $E\simeq 1.3$ keV, which might be third or fourth harmonic
of the cyclotron energy, $E_{c,e}^\infty\approx 0.33$ or 0.26 keV, respectively\footnote{Halpern 
\& Gotthelf (2010) do not mention this feature. We note that
it is seen in the EPIC pn and MOS spectra,
which supports the reality of this feature despite its low 
statistical significance in each of the spectra.}. To confirm the feature
and estimate the cyclotron energy and gravitational redshift, 
a deeper observation of this CCO is required.

For the other
pulsating CCO, J0822--4300 in the Puppis A SNR, the period derivative has not yet been measured,
and only an upper limit on the magnetic field, $B_{\rm sd}<9.8\times 10^{11}$ G,
 has been estimated by Gotthelf \& Halpern (2009).
Fitting the spectrum with a double blackbody model, these authors found an emission feature
around 0.8 keV at some phases. 
To explain the fact that the feature is seen in emission, one has to assume an energy source in the upper layers of
the atmosphere, such as the heat released by accreting matter, which is not included in
our models. 
However, 
it seems that 
the spectral structure at these energies (see Figure 3 in 
Gotthelf \& Halpern 2009) 
can also be interpreted as 
an absorption feature, perhaps a cyclotron line, centered at $\approx 0.9$ keV,
which of course would imply a different continuum. 
To examine this interpretation and look for quantum oscillations, the CCO's phase-resolved
 spectra should be
fitted with our atmosphere models.

Another class of NSs
with possible 
quantum oscillations in their spectra
are 
radio pulsars with 
magnetic fields $10^{10}$--$10^{11}$ G.
There are 62 pulsars with such fields in the ATNF Pulsar Catalogue\footnote{See {\tt http://www.atnf.csiro.au/research/pulsar/psrcat} and  Manchester et al.\ (2005).}, but none of them has been observed
in X-rays. As these pulsars are old
(the youngest one, PSR J1810--1820
with $B_{\rm sd}=4.3\times 10^{10}$ G, has
the spin-down age $\tau_{\rm sd}=P/2\dot{P}=47$ Myr),
the bulk of 
NS surface 
is 
too cold to be seen in X-rays, 
but their polar caps
can be heated up to 
$\sim 1$--3 MK
 by relativistic particles and $\gamma$-rays generated
in the pulsar magnetosphere (see, e.g., Zavlin \& Pavlov 2004).
Observing the quantum oscillations in the thermal spectra of polar caps is, however, a challenging
task because the luminosity of this component is low and 
it is strongly absorbed by the ISM for many of the pulsars.

%In principle, similar quantum oscillations could also arise due to 
%quantization of the {\em proton}
%motion in a strong magnetic field, if the proton quantization parameter,
% $b_p = E_{c,p}/kT = 0.73\, (B/10^{13}\,{\rm G})(T/1\,{\rm MK})^{-1}$, is sufficiently
%large. 
%We cannot rule out the possibility that the harmonically spaced absorption
%features detected in DINSs (see Introduction) can be explained by this effect,
%but accurate calculations of opacities near the proton cyclotron resonances
%are required to check this hypothesis\footnote{
%Potekhin \& Chabrier (2003) found that the peaks of opacity at the proton cyclotron
%harmonics are very weak, but Sawyer (2007) called their results into question.}.

In conclusion, we should mention that 
the opacities by PP76, used in our atmosphere models,
become inaccurate close to the 
centers of the quantum peaks, especially at the core
of the fundamental cyclotron line. A more accurate consideration, which 
would include corrections to
the Born approximation and 
the collective (high-density) effects
(e.g., Sawyer 2007),
and take into account the Doppler broadening of the fundamental
resonance (Pavlov et al.\ 1980a), 
should broaden the spectral features and decrease the number of observable ones.
Also, 
it would be worthwhile to investigate the
role of the relativistic corrections to the opacities (including
the resonances at the cyclotron harmonics).
Finally, for the comparison with the observational data, it would be useful to
construct atmosphere models with magnetic fields inclined to the normal to the surface
and integrate the 
emission over (a fraction of) the NS surface. We plan to consider these problems in future
works.

\acknowledgments

VS thanks DFG for financial support (grant 
SFB/Transregio 7 ``Gravitational Wave Astronomy''). 
The work by GGP was partially supported by NASA grant NNX09AC84G.
We thank Raymond Sawyer for useful discussions.

%\clearpage

\end{document}